\documentclass[10pt,preprint2]{aastex}
\usepackage{color}
\usepackage{url}
\usepackage{bm}

\shorttitle{Applicability of a Flare Trigger Model}
\shortauthors{Bamba et al.}

\begin{document}

\title{Evaluation of Applicability of a Flare Trigger Model\\based on Comparison of Geometric Structures}

\author{Yumi Bamba}
\affil{Institute of Space and Astronautical Science (ISAS)/Japan Aerospace Exploration Agency (JAXA)\\ 3-1-1 Yoshinodai, Chuo-ku, Sagamihara, Kanagawa 252-5210, Japan}
\email{y-bamba@nagoya-u.jp}

\author{Kanya Kusano}
\affil{Institute for Space-Earth Environmental Research (ISEE)/Nagoya University\\ Furo-cho, Chikusa-ku, Nagoya, Aichi 464-8601, Japan}

%%%ABSTRACT%%%
\begin{abstract} %less than 250 words

The triggering mechanism(s) and critical condition(s) of solar flares are still not completely clarified, although various studies have attempted to elucidate them. We have also proposed a theoretical flare-trigger model based on MHD simulations \citep{kusano12}, in which two types of small-scale bipole field, the so-called Opposite Polarity (OP) and Reversed Shear (RS) types of field, can trigger flares. In this study, we evaluated the applicability of our flare-trigger model to observation of 32 flares that were observed by the Solar Dynamics Observatory (SDO), by focusing on geometrical structures. We classified the events into six types, including the OP and RS types, based on photospheric magnetic field configuration, presence of precursor brightenings, and shape of the initial flare ribbons. As a result, we found that approximately 30\% of the flares were consistent with our flare-trigger model, and the number of RS type triggered flares is larger than that of the OP type. We found none of the sampled events contradicts our flare model, although we cannot clearly determine the trigger mechanism of 70\% of the flares in this study. We carefully investigated the applicability of our flare-trigger model and the possibility that other models can explain the other 70\% of the events. Consequently, we concluded that our flare-trigger model has certainly proposed important conditions for flare-triggering.

\end{abstract}


%%%%%%%%%%%%%%%%%%%%%%%%%%%%%%%%%%%%%%%%%%%%%%%%%%%%%%%%%%%%%%%%%%

%%%Sec 1%%%
\section{Introduction} \label{sec:intro}

A solar flare is a sudden release of magnetic energy in the solar corona, and it is widely accepted that flare occurrence is related to topological changes of the magnetic field such as caused by magnetic reconnection (cf. CSHKP model, \citet{carmichael64, sturrock66, hirayama74, kopppneuman76}).
Two major physical models of solar eruptions including flares and coronal mass ejections (CMEs) are proposed so far: ``ideal magnetohydrodynamics (MHD) models'' and ``magnetic reconnection models''.
They emphasize different aspects of the mechanism of solar eruptions.
The ideal MHD models point out that some ideal MHD instabilities cause the onset of solar eruptions.
For instance, the torus instability \citep[e.g.][]{bateman78, kliemtorok06, demoulin10, kliem14} or helical kink instability \citep[e.g.][]{gerrard01, torok04, fangibson03, torokkliem05} have been proposed as the mechanism of solar eruptions.
The torus instability is determined by the decay rate of the external poloidal field $B_{ex}$, and the critical condition is defined by the decay index ${n = -d({ln} B_{ex})/d({ln} R)}$.
Recently, the critical range for onset of the torus instability is proposed as $n\sim1.3-1.5$ \citep{zuccarello15}. 
Some observational and computational studies using non-linear force-free field (NLFFF) \citep[e.g.][]{cheng11, kliem13, savcheva15} modeling support the torus instability although the torus instability required an external agent that lifts the flux rope to an unstable height range.
The helical kink instability model proposes that solar eruptions are caused by growth of the helical kink instability.
The helical kink instability can grow when magnetic twist becomes strong enough, and it can grow even if a twisted flux rope does not reach the critical height of the torus instability.
\citet{hassanin16} showed a nice comparison between observations and an MHD simulation for a confined M-class flare.
\citet{liu16} also showed that there is consistency between the helical kink instability mechanisms and sequential confined flares from a combination of observations and NLFFF modeling.
However, the helical kink instability requires higher twist in a flux rope and observations do not always show such a high twist.

The ideal MHD models postulate that a flux rope exists prior to eruption as mentioned above.
On the other hand, magnetic reconnection models note that some kind of magnetic reconnection causes the onset of a solar eruption.
Two kinds of reconnection models have been proposed so far.
One is the ``magnetic breakout model'' \citep{Antiochos99, karpen12} and the other is the ``tether-cutting model'' \citep{moore01, moore06}.
The magnetic breakout model considers a multi-flux topology that consists of multiple flux systems with a coronal null point \citep[e.g.][]{priest96, sun12}.
Sheared magnetic arcades can be created through many means such as slow photospheric shear. This leads to storage of free energy. Breakout reconnection can then remove the overlying magnetic field, and the core magnetic flux is free to erupt.
\citet{reva16} showed breakout reconnection associated with a CME using observations from TESIS EUV telescope.
The tether-cutting model proposes that magnetic reconnection in the core of a sheared magnetic arcade creates a flux rope and causes the eruption.
\citet{liu13} showed evidence of tether-cutting reconnection using a NLFFF model using solar observational data of a flare productive AR.
The breakout and tether-cutting models are clearly reviewed from an observational perspective in \citet{schmieder15}.

There are various studies trying to find out what observable parameter(s) determine flare and solar eruption onset condition(s), while various models are proposed from computational studies and have been validated by observations, as mentioned above.
One such model is flux cancellation \citep{vanBallegooijen89, aulanier10}, whose basic concept is that tether-cutting reconnection forms a flux rope and it partially cancels magnetic flux in the core of a sheared arcade causing the MHD instability to be triggered.
\citet{Green11} clearly showed flux cancelation and associated flux rope formation in the photoshere prior to an eruption.
Another model is the emerging flux model proposed by \citet{chenshibata00}.
They considered two cases based on the positional relationship between emerging flux and a flux rope.
If emerging flux appears below a flux rope (i.e. on the polarity inversion line: PIL) and partial magnetic cancellation occurs, local magnetic pressure is decreased and the pressure gradient causes upflow that pushes the flux rope upward to erupt.
On the other hand, if a flux rope appears beside the overlying field (i.e. away from the PIL), the overlying field that prohibits the flux rope from eruption is weakened by magnetic reconnection with the emerging flux, and the flux rope may erupt.
Many other studies proposed various magnetic properties, that seem to relate to the onset of solar eruptions, such as length of highly sheared PILs \citep{hagyard84b}, magnetic flux close to high gradient PILs \citep{schrijver07}, free magnetic energy based on presence of strong gradient PILs \citep{falconer08}, area and total magnetic flux in an active region (AR) \citep{higgins11}, and total unsigned current helicity over an AR \citep{bobra15}, etc.
However, it is still unclear what triggers the MHD instability or magnetic reconnection that leads to an eruption.

Although each model proposed a solar eruption scenario, it is likely that the eruption onset results from the feedback and interaction of different processes.
For instance, \citet{hagyard84a} suggested that an overall instability driving energy release results from the positive feedback between reconnection and eruption of the sheared field.
\citet{kusano12} (henceforth ``K12'') numerically demonstrated the feedback model, in which positive feedback between an ideal MHD instability and magnetic reconnection causes explosive growth of solar eruptions.
Their basic idea is that ``internal magnetic reconnection'' between different spatial-scale magnetic fields (between a large-scale sheared field and a small-scale bipole field) forms a double-arc twisted loop and triggers reconnection among the large-scale sheared fields by destabilizing the double-arc twisted loop.
In fact, \citet{ishiguro17} recently showed that the double-arc twisted loop becomes unstable if the parameter $\kappa$, given by the product of magnetic twist and the fraction of reconnected flux, exceeds a threshold.
Their model is consistent with the tether-cutting reconnection scenario and well explains the physical process from the theoretical point of view.
K12 proposed that the feedback interaction can be triggered by reconnection between magnetic fields of different spatial-scale if two geometrical parameters; sheared angle $\theta_{0}$ of large-scale field and azimuth $\varphi_{e}$ of a small-scale bipole field, satisfy some condition.
We briefly review the K12 model later in Section~\ref{sec:KB12} and Figure~\ref{fig:setting}.
They surveyed combinations of $\theta_{0}$ and $\varphi_{e}$ and found that there are two specific cases; the Opposite Polarity (OP) and Reversed Shear (RS) fields, which can trigger a flare.
The essence of the K12 model is that a positive feedback interaction of an ideal MHD instability and flare reconnection can be triggered by the two types of small-scale ``trigger fields''.
\citet{bamba13} developed a way to measure the angles $\theta_{0}$ and $\varphi_{e}$ using the photospheric magnetic field and chromospheric brightenings observed by the Hinode satellite \citep{kosugi07}.
They found that either the OP or RS type magnetic field condition was satisfied prior to several major flares.

In this study,  we aim to evaluate the consistency between the flare-trigger condition of K12 and observations, focusing on geometrical structures.
For that purpose, we investigate more flares than \citet{bamba13} since only a small number of flares were studied in their papers.
Specifically, we determine what fraction of the flares are consistent with the OP or RS types and whether a flare occurred with some condition other than OP or RS field.
We increased the number of events analyzed to 32.
This was possible due to the large field-of-view (FOV) of the Solar Dynamics Observatory (SDO; \citet{pesnell12}).

The paper is organized in the following manner.
We first review the theoretical model of K12, which we focused on this study, and describe the data analysis methods and event selection criteria in Section~\ref{sec:data_analysis}.
Then we classified the events into six different types, including the OP and RS types, based on the analysis, and describe distinctive features for each type, in Section~\ref{sec:results}.
In Section~\ref{sec:discussion}, we discuss other possibilities than K12's model for triggering some specific types of events which we classified in this study, and discuss caveats to the analysis identifying the flare-triggering site.
We finally summarize the conclusions of the present study in Section~\ref{sec:summary}.

%%%Sec 2%%%
\section{Data Analysis and Event Classification} \label{sec:data_analysis}

\subsection{Synopsis of the Theoretical Model} \label{sec:KB12}

We briefly review the theoretical model proposed by K12.
The model simplifies the magnetic field structure of an AR and quantitatively surveyed magnetic configurations which can produce solar eruptions (flares).
Figure~\ref{fig:setting}(a) shows their simulation setup.
They imposed a small bipole field onto the PIL of the large-scale bipole field representing an active region.
The internal magnetic reconnection can occur between the sheared field (red arcade) and the small-bipole field (blue arcade), and it precedes flare reconnection among the large-scale sheared fields (red arcades) distributed along the PIL if some condition is satisfied.
The surveyed parameters, which describe whether flares occur or not, are $\theta_{0}$ and $\varphi_{e}$; defined as illustrated in both panels (a, b).
$\theta_{0}$ is the angle of the magnetic field relative to the potential field and it increases counter-clockwise in the range of 0$^{\circ}$-90$^{\circ}$.
$\varphi_{e}$ is the counter-clockwise rotation angle of the small-scale bipole field relative to the large-scale potential field of in the range of 0$^{\circ}$-360$^{\circ}$.
The parameters $\theta_{0}$ and $\varphi_{e}$ correspond to the magnetic twist and complexity of the magnetic field, respectively.
The combination of ($\theta_{0}$, $\varphi_{e}$) characterizes the flare-trigger conditions, and either the OP or RS type bipole field can trigger the positive feedback process of an MHD instability and magnetic reconnection, leading to flares (and sometimes CMEs).
The pre-existing sheared fields connect via the OP type field and form a double-arc twisted flux rope, then an eruptive MHD instability is triggered and the instability causes flare reconnection, i.e., ``eruption-induced reconnection'' starts.
If the RS type of field exists, a part of the pre-existing sheared field is cancelled by reconnection with the RS field in the core region, and the sheared field collapses inwards because magnetic pressure is decreased.
Then flare reconnection is triggered and the twisted flux rope erupts.
It can be described as a ``reconnection-induced eruption'' process since the erupting flux rope is formed after the flare reconnection.

Several observables which are relevant to the flare-trigger condition, have been derived from the simulations, and are summarized below.

\renewcommand{\theenumi}{\Roman{enumi}}
\begin{enumerate}

	\item[I.] The magnetic shear angle $\theta_{0}$ and the angle structure of the PIL disturbance at small scales $\varphi_{e}$ is observed to be consist with the conditions of either the OP or RS type.
	
	\item[I\hspace{-.1em}I.] Precursor brightenings should be seen near the PIL where the OP or RS configuration is satisfied. This represents the internal reconnection in the lower atmosphere. Hence it is inferred that precursor brightenings can be observed in any emission line which is formed in the lower atmosphere, for example, the chromosphere.
	
	\item[I\hspace{-.1em}I\hspace{-.1em}I.] The flare ribbons, which appear in the initial flare phase, should have a clearly sheared configuration as illustrated by the thick yellow lines in Figure~\ref{fig:setting}. It results from a theoretical prediction that the flare reconnection can be reinforced by an instability which is driven by large-scale (non-potential) fields (such as the red arcades in Figure~\ref{fig:setting}).
	
\end{enumerate}

A sheared two-ribbon structure and precursor brightening on a highly sheared PIL is common in many other models such as flux cancellation, flux emergence, any tether-cutting reconnection, and also magnetic breakout models.
The key feature of the K12 model is that the precursor brightening should appear near the local PIL where the OP or RS conditions are satisfied.
It is due to the fact that precursor brightenings may represent local heating caused by the internal reconnection between the pre-existing sheared field and trigger field.
Moreover, \citet{bamba13, bamba14} reported that precursor brightenings were observed in the lower atmosphere from the upper photosphere to the transition region, rather than in the corona.
This is consistent with the K12 simulation in which the OP or RS flux is injected from the bottom boundary into the corona.
Therefore, in this study, we use the above three features (geometrical structure of magnetic field; precursor brightening in the chromosphere; and initial flare ribbons) to test the K12 model.

\subsection{Data Description} \label{sec:data}

We used data obtained by the Helioseismic and Magnetic Imager (HMI, \citet{schou12}) and the Atmospheric Imaging Assembly (AIA, \citet{lemen12}), both are onboard the SDO satellite.
These instruments observe the full disk of the Sun with a large FOV of $2000^{\prime\prime} \times 2000^{\prime\prime}$.
HMI observes the polarization states Stokes-I, Q, U, and V in the photospheric Fe I line (6173{\AA}) with a spatial resolution of $1^{\prime\prime}$.
In this study, we used the HMI level 1.5 line-of-sight (LOS) magnetograms ({\tt hmi.M\_45s} series) and the Spaceweather HMI Active Region Patch (SHARP, {\tt hmi.sharp\_cea\_720s}, \citet{bobra14}) series, that contains vertical and horizontal components of the photospheric magnetic field.
The $180^{\circ}$ ambiguity is resolved and magnetic field vectors are remapped to the Lambert Cylindrical Equal-Area (CEA) projection for SHARP data.
Data cadences are 45 sec. for the LOS magnetograms and 720 sec. for SHARP data, respectively.
AIA observes solar atmosphere in ten EUV and UV channels, but in this study, we only used the AIA 1600 {\AA} (continuum and C IV line) images that are sensitive to emission from the upper photosphere and transition region (log T $\sim 5.0$).
The spatial resolution is $1.5^{\prime\prime}$ and the cadence is 24 sec. for the 1600 {\AA} data.

\subsection{Analysis Procedure} \label{sec:method}

We followed the analysis method developed in \citet{bamba13}.
The analysis method was developed for Hinode data, and \citet{bamba14} examined the applicability of the techniques to SDO data.
Here we summarize the essence of the procedure (see details in \citet{bamba13, bamba14}).

\renewcommand{\theenumi}{\arabic{enumi}}
\begin{enumerate}

	\item We calibrated the HMI level 1.5 LOS magnetograms and AIA level 1.0 images to remove spatial fluctuations (spacecraft jitter) and to resample the HMI and AIA images to the same size.
	
	\item We superimposed the PILs and strong brightening contours seen in AIA images onto the HMI LOS magnetograms. We define the flare-triggering site as a region located where the center of the initially sheared flare ribbons and precursor brightening were seen. In other words, if there is no precursor brightening between the initially sheared flare ribbons, we are not able to define the flare-triggering site for the event.
		
	\item We measured the magnetic shear angle $\theta_{0}$ along the flaring PIL and azimuth $\varphi_{e}$ at the flare-triggering site. $\theta_{0}$ was measured as the angle of the initially sheared flare ribbons relative to direction $\bm{N}$, perpendicular to the averaged PIL and representing the direction of the potential field. The averaged PIL used to determine $\bm{N}$ was derived from a low-pass filtering of the LOS magnetic field through a Fourier transformation. The angle of the initially sheared flare ribbons was determined by the averaged angle of the sheared ribbons. We therefore are able to measure the angle only for the events that show two clear flare ribbons in the initial phase. $\varphi_{e}$ is the angle perpendicular to the local PIL in the flare-triggering site relative to $\bm{N}$.
		
\end{enumerate}

\subsection{Event Selection and Classification Criteria} \label{sec:criteria}

Following to \citet{bamba13}, we selected events by the criteria summarized below.

\begin{itemize}
	\item GOES class was larger than M5.0.
	\item Event occurred in the period of 2010 February 11 to 2014 February 28.
	\item The SDO/HMI and AIA 1600 {\AA} data covered a period of six hours before and after the flare onset for each event.
	\item Flaring site was located within $\pm~750^{\prime\prime}$ from the solar disk center.
\end{itemize}

\noindent Then we sampled 32 flare events, as summarized in Table~\ref{table:event_list}, and classified the events into six types by evaluating whether the important features I, I\hspace{-.1em}I, and I\hspace{-.1em}I\hspace{-.1em}I of Section~\ref{sec:KB12} were observed.
The classification procedure and criteria are summarized in Figure~\ref{fig:chart}, and we labeled each type of event as below.
The quantitative values of the angles ($\theta_{0}$, $\varphi_{e}$) stated below were uniquely defined in the present study based on the simulations of K12 for the event classification.
Note that we only measured the angles for the events which showed two clear sheared ribbons and precursor brightenings at a single position, hence we did not measure the angles for any events classified as the Multiple Trigger Candidates type.

\begin{description}

	\item[Opposite Polarity (OP) type]\mbox{}\\
	The event satisfies the ($\theta_{0}$, $\varphi_{e}$) condition of the OP type in Figure 2 of K12. In particular, the shear angle and azimuth should satisfy $0^{\circ} \leq \theta_{0} \leq 90^{\circ}$ and $124^{\circ} \leq \varphi_{e} \leq 225^{\circ}$ at the region/timing where/when the last precursor brightening was seen. Qualitatively, the magnetic field in the small-bipole region has an opposite polarity pattern relative to that averaged over the whole AR.
	
	\item[Reversed Shear (RS) type]\mbox{}\\
	The event satisfies the ($\theta_{0}$, $\varphi_{e}$) condition of the RS type in Figure 2 of K12. In particular, the shear angle and azimuth should satisfy $40^{\circ} \leq \theta_{0}$ and $225^{\circ} \leq \varphi_{e} \leq 335^{\circ}$ at the region/timing where/when the last precursor brightening was seen. The RS type has a $\theta_{0}$ range of approximately $40-90^{\circ}$. The local magnetic shear in the small-scale bipole region is towards the opposite direction of the global magnetic shear of the AR. 
		
	\item[Contradict-K12 type]\mbox{}\\
	The event occurs ``no-flare'' ($\theta_{0}$, $\varphi_{e}$) condition in Figure 2 of K12. In particular, $0^{\circ} \leq \varphi_{e} \leq 120^{\circ}$ and $250^{\circ} \leq \varphi_{e}$ with a small $\theta_{0}$ value at the region/timing where/when the last precursor brightening was seen.
	
	\item[Multiple Trigger Candidates type]\mbox{}\\
	The event clearly shows two sheared ribbons (similar to the thick yellow lines in Figure~\ref{fig:setting}) in the initial flare phase. However, there are multiple small-scale bipole fields on which precursor brightenings are observed on the local PIL before the flare onset. In other words, there are multiple small-scale bipole fields which show the important features I\hspace{-1pt}I. and I\hspace{-1pt}I\hspace{-1pt}I. of Section~\ref{sec:KB12}, before the flare onset. 
		
	\item[No-precursor Brightening type]\mbox{}\\
	The event clearly shows two sheared ribbons in the initial flare phase. However, it does not show any precursor brightening over the local PIL of the small-scale bipole field located at the center of the two ribbons, that exists for at least two observation frames (approximately at least over a period of 1.5 min.) in AIA 1600 {\AA} images from six hours before to the flare onset time. In brief, the event shows only the important structure I\hspace{-1pt}I\hspace{-1pt}I. of Section~\ref{sec:KB12}.
	
	\item[Complicated Ribbon Type]\mbox{}\\
	The event shows complicated initial flare ribbons which are nothing like the thick yellow lines in Figure~\ref{fig:setting}, i.e., they are much different from feature I\hspace{-1pt}I\hspace{-1pt}I suggested by the simulations. More specifically, the flare ribbons seen in the AIA 1600 {\AA} images overplotted on the HMI LOS magnetogram are not able to distinguish its positive and negative polarities during the period from 5 minutes before flare onset to peak time. This type includes cases where flare ribbons appear in parallel (not sheared like Figure~\ref{fig:setting}).
	
\end{description}

%%%Sec 3%%%
\section{Results} \label{sec:results}

The selected flare events and classification results are listed in Table~\ref{table:event_list}, and Table~\ref{table:list_summary} is a summary of the classification.
First, we did not find any events of the Contradict-K12 type.
It means that there was no flare event which was investigated in this study that occurred under ``no-flare'' combinations of $(\theta_{0}, \varphi_{e})$.
It suggests that there was no event that was clearly inconsistent with the flare-trigger conditions proposed by K12, at least among the events investigated in the present study.

Second, six of the 32 events satisfied the RS type condition.
Events No. 1 and 2 are the same RS type events examined in previous studies (\citet{bamba13, bamba14}).
Figure~\ref{fig:sample}(a-1)-(a-3) shows an example of an RS type event: the X5.4 flare that occurred on 2012 March 7 in AR NOAA 11429.
The white/black background indicates the positive/negative polarity of the LOS magnetic field.
The green lines and red contours outline the PILs (0 G lines) of the LOS magnetic field and strong emission (2000 DN) in AIA 1600 {\AA}, respectively.
The image shows flare ribbons that initially form two clearly sheared ribbon structures on both sides of the PIL, as outlined by black and white broken lines.
The yellow arrow indicates the flare-triggering site, and precursor brightenings were intermittently seen in the region, as reported in \citet{bamba13, bamba14}.
We measured the magnetic shear angle $\theta_{0}$ and azimuth $\varphi_{e}$ around the flare-triggering site, and the angles were ($\theta_{0}, \varphi_{e})\sim(108^{\circ}, 313^{\circ}$) in the case of the X5.4 flare as summarized in Table~\ref{table:event_list}.
In contrast, there was no OP type event in our selected sample.
Note that  \citet{bamba13} showed two examples of OP type events that occurred in 2006, that are out of the time range of this study because the events occurred before the SDO launch.

Third, four of the 32 events were classified as Multiple Trigger Candidates type.
As an example, the M6.0 flare that occurred on 2011 August 3 in AR NOAA 11261 is shown in Figure~\ref{fig:sample}(b-1)-(b-3).
These events showed the important features I. and I\hspace{-1pt}I.
However, there were multiple candidates for precursor brightenings at different locations, and a single flare-triggering site could not be identified.
In the case of the RS type events, a precursor brightening seen at a single location between the initial two ribbons.
However, in the Multiple Trigger Candidates type events, precursor brightenings were intermittently seen at several points on the PIL that was in between the two sheared ribbons.
Hence, it was difficult to conclude that a single PIL was the flare-triggering site.
Moreover, magnetic field structures in the precursor brightening sites were very small spatially in some cases, less than $2^{\prime\prime}$, that is close to the $1^{\prime\prime}$ spatial resolution of HMI.
Especially, azimuth $\varphi_{e}$, which is measured at the point of the precursor brightening site, is highly sensitive to the spatial configuration (i.e. the shape of the PIL) of the site.
In this study, we chose the flare-triggering sites to be the location where there were precursor brightenings, and for these cases, it was extremely difficult to measure $\varphi_{e}$ at the correct location.
Therefore, in this study, we did not measure the angles for these four Multiple Trigger Candidates type events.
Note that the lifetime and spatial size of the precursor brightenings seen in the RS and Multiple Trigger Candidates (and OP) types are different between different events.
\citet{bamba13} suggested that the critical magnetic flux differs among events because the critical perturbation amplitude needed to trigger the instability depends on how the system is unstable.
Related to this, the spatial size and duration of the heating caused by internal reconnection within the trigger field represented by precursor brightenings is different among the events.
In addition, empirically from this and other studies \citep{bamba13, bamba14, Palacios15, bamba17a, bamba17b, Wang17, Woods17}, the duration varies from several hours to a few seconds and the spatial size were from a few to dozens of arcseconds.

Another noticeable thing is that 22 events were classified as No-precursor Brightening or Complicated Ribbon type.
In other words, approximately 70\% of the selected events did not show any clear features that were predicted by K12's MHD simulation such as sheared two-ribbon structure or precursor brightenings.
Panels (c-1)-(c-3) show an example of a No-precursor Brightening type event: the M6.5 flare that occurred on 2013 April 11 in AR NOAA 118719.
The 11 events that are classified as No-precursor Brightening type clearly showed a sheared two-ribbon structure in the initial flare phase as suggested by the simulation.
However, no brightening was seen on the PIL between the two sheared ribbons, in at least two observation frames (approximately at least over a period of 1.5 min.), in AIA 1600 {\AA} images, from six hours before to the flare onset time.
Many of the No-precursor Brightening type events showed similar behavior: that the initial flare ribbons did not widely propagate.
The flare ribbons suddenly appeared as bright points, and the intensity became enhanced at almost the same location within a few minutes.
Panels (d-1)-(d-3) show an example of the Complicated Ribbon type: the M9.3 flare that occurred on 2013 October 24 in AR NOAA 11877.
Initially the flare ribbons had a complex shape or sometimes looked almost like a single ribbon.

Here, we briefly summarize the classification results.
\begin{itemize}
	\item There was no Contradict-K12 type event that satisfied the ``no-flare'' condition of K12.
	\item 30\% of the events, including six RS type events and four Multiple Trigger Candidates type, were consistent with K12.
	\item Approximately 70\% of the events (22 of the analyzed 32 events) were classified as either the No-precursor Brightening or Complicated Ribbon types that did not clearly show the key features suggested by K12.
\end{itemize}

%%%Sec 4%%%
\section{Discussion} \label{sec:discussion}

The result that 70\% of the events did not show the key features suggested by the MHD simulation of K12 is interesting.
Therefore, we here discuss whether the events classified as Complicated Ribbon and No-precursor Brightening type could be explained by K12's model or other models.

\subsection{Complicated Ribbon type} \label{sec:dis_CR}

It is not surprising that many of the flares showed complicated flare ribbons because the coronal magnetic fields are naturally intertwined. 
Figure~\ref{fig:d_11515} shows a time series analysis of the M5.6 flare (No. 17 listed in Table~\ref{table:event_list}) that occurred on 2012 July 2 in AR NOAA 11515, which was classified as a Complicated Ribbon type.
The M5.6 flare occurred in a weak negative polarity region between the leading sunspot (LS) and satellite spot (SS).
The red contour in panel (e) shows the initial flare ribbon of the M5.6 flare, but it is very different from the sheared two-ribbon structure which is suggested by K12's simulation (illustrated by the yellow thick line in Figure~\ref{fig:setting}).
The M5.6 flare followed a filament eruption and the preceding C2.9 flare whose flare ribbons are seen in Figure~\ref{fig:d_11515}(d) in the same region, as \citet{louis14} reported.
The filament was rooted in the rear part of the LS and weak negative region as illustrated in Figure~\ref{fig:d_11515}(a, b) by blue arcs.

We found a tiny brightening near the southern root of the filament (the region surrounded by the yellow square in Figure~\ref{fig:d_11515}(a, b)) over a small wedge-like structure, as can be seen in the enlarged image in panel (a), and it started from 10:10 UT on 2012 July 2.
The brightening region gradually extended (as seen in panel (b)), and the C2.9 flare occurred at 10:33 UT (panel (c)).
The flare ribbons of the C2.9 event were also complicated as seen in panel (d).
The M5.6 flare started at 10:43 UT (panel (e)) from a region slightly to the northwest of the C-class flaring region.
Then two clear ribbons (PR and NR) appeared as sheared ribbons in the latter phase (panel (f)).
We measured the angles $\theta_{0}$ and $\varphi_{e}$ before the C2.9 flare and eruption onset, assuming that the large-scale magnetic field structure around the SS had not drastically changed.
The result of the measurement was $(\theta_{0}, \varphi_{e})\sim(126^{\circ}, 351^{\circ})$, and it is consistent with the RS type condition.
From the result, we were able to conclude that the M5.6 flare was triggered by the RS type magnetic field structure in a three-step process, including the C2.9 flare and the filament eruption.

\citet{bamba17a} also identified the flare-triggering site for an X1.0 flare, which showed three complicated flare ribbons in the initial phase and which can be classified as Complicated Ribbon type, through a detailed analysis.
They suggested that precise analyses of spatial and temporal relationships between magnetic field structures and overlying structures such as filaments and precursor brightenings are effective for identifying the flare-triggering site for the Complicated Ribbon type event.
Further, \citet{bamba17b} confirmed by spectroscopic observation that precursor brightenings can be a marker of magnetic reconnection in the lower atmosphere, such as the internal magnetic reconnection proposed by K12.
We hence investigated whether precursor brightenings were seen on a PIL in the region where flare ribbons appeared with regards to the Complicated Ribbon type events.
Table~\ref{table:list_typeD} shows a list of the Complicated Ribbon type events and whether precursor brightenings exist or not.
Brightenings on a PIL were observed in eight events out of the 11 Complicated Ribbon type events.
Accordingly, we propose that it is possible that the flare-triggering site for the eight events may be found through detailed analysis, and we need precise analysis for each individual event.

On the other hand, we further should consider the relationship between the Complicated Ribbon type events and a flare-trigger by magnetic reconnection at a coronal null point \citep{Longcope05, Titov07, Titov09}.
The flare ribbons of the above M5.6 flare (seen in Figure~\ref{fig:d_11515}(f)) look like the circular-shaped ribbons such as studied by for example, \citet{Masson09}.
Complex circular-shaped or scattered flare ribbons distributed from a coronal null point origin may be possible in this case.

\subsection{No-precursor Brightening type} \label{sec:dis_NB}

With regard to the events which did not cause precursor brightenings such as the No-precursor Brightening type and three of the Complicated Ribbon type events, the possibility that the flares were triggered by a different physical process(es) from that proposed by K12 cannot also be ruled out.
K12 proposed that major flares should be preceded by internal magnetic reconnection between the flare-triggering flux of the OP or RS types and pre-existing sheared magnetic field.
They treated precursor brightenings in the lower atmosphere on the flare-triggering site as a proxy of the internal reconnection.
Therefore, we still can consider other physical process(es).
For instance, if a long twisted flux rope was strapped down by overlying magnetic field and the trapping field was weakened by some cause such as emerging flux (case-B scenario of \citet{chenshibata00}), then in this case, precursor brightenings may be seen at a foot point of the trapping field, which may be away from the flaring site.
We should confirm theexistence of a long twisted flux rope, and investigate how the flux rope was formed during the AR evolution, using observational data at high temperature, such as taken by the Hinode/X-ray telescope.
Otherwise, gradual magnetic flux cancellation at a foot point of the overlying field may also weaken the trapping field \citep{vanBallegooijen89, Zhang01, Green11}.
Precursor brightenings may be observed if the flux cancellation was caused by magnetic reconnection between opposite polarity fluxes.
However, it is conceivable that no brightening may be observed in the case of flux cancellation by sinking of small magnetic fluxes in the photosphere.
Another possibility for no-precursor brightening event is that a magnetic flux tube had formed in advance by some cause such as a preceding flare or photospheric motion and it became unstable to the ideal MHD instabilities of torus, kink and double-arc modes \citep{kliemtorok06, torokkliem05, ishiguro17}.

We note that there may be a possibility that precursor brightenings can be seen in a different wavelength from AIA 1600 {\AA}, which was used in the present study and which is sensitive to emission from the upper photosphere and lower transition region (log T $\sim 5.0$).
Especially, in the case of the RS type, magnetic shear cancellation by internal reconnection could occur at any altitude while the internal reconnection should occur in the lower atmosphere in the OP type case.
This is because the altitude of the internal reconnection in the RS type case depends on the spatial size of the pre-existing sheared field and flare-trigger field.
Therefore, it might be likely that we cannot detect internal reconnection of the RS type by emission in 1600 {\AA}, if the internal reconnection proceeds in the coronal region.
We can also consider the breakout trigger scenario \citep{Antiochos99}, in which magnetic reconnection occurs at much higher altitude than focused on in this study.
Precursor brightenings may be observed in higher temperature lines than AIA 1600 {\AA} as in the RS type case event.
Another possibility is that the size of the internal reconnection region working as a trigger was too small to be observed.
Recently, \citet{ishiguro17} developed a theoretical model of the new Double-Arc Instability, and according to that model they predict that the size of the triggering reconnection depends on the strength of the magnetic twist.
If magnetic twist is high enough, even tiny reconnection can trigger the Double-Arc Instability and work as a flare trigger.
Therefore, it is necessary to distinguish which types of flares are caused by which physical process, i.e., which types of flares are explained by which model, by more precisely analyzing each of the events using multiple wavelengths data.

\subsection{Multiple Trigger Candidates type} \label{sec:dis_MTR}

The Multiple Trigger Candidates type events, in which there were several flare-triggering site candidates, is still understandable by the flare-trigger model of K12.
It is not necessary to restrict the number of flare-triggering sites for some cases.
\citet{bamba13, bamba14} suggested that not only the geometrical conditions $\theta_{0}$ and $\varphi_{e}$ but also the total magnetic flux in the flare-triggering site and/or its temporal evolution contribute to the critical conditions for flare occurrence.
They also suggested that the critical magnetic flux required to trigger an instability depends on the proximity of the system to an unstable state.
Hence the combined contribution of more than one small-scale flux, that satisfies either the OP and/or RS type condition, is also conceivable as a flare-trigger.
\citet{bamba17b} analyzed a region which includes two RS type bipole structures, and studied the sequential process from the internal reconnection, which was represented by precursor brightenings, to destabilization of the large-scale system in the AR, using spectroscopic data.
However, they were not able to determine the detailed process by which the two flare-triggering sites contributed to the flare occurrence because the data they employed only covered the odd RS type region.
Thus, we need more analyses to clarify the combination effect of multiple and different types of trigger regions.
These No-precursor Brightening and Multiple Trigger Candidates type events are future topics of discussion.

\subsection{Difference of the Incidence rate of OP and RS types} \label{sec:incidence}

Another interesting topic for discussion is the result that the OP type events were rare compared to the RS types, at least in this study.
Some other studies found events which were suggested to be triggered by the OP type small magnetic structure \citep{Palacios15, Wang17, Woods17}.
However, so far we found only two OP type flare events (already shown by \citet{bamba13} by the same method of this study) whereas six events were classified as the RS type in this study.
We consider that the difference in incidence between the two types results from a difference of flexibility in the geometrical conditions, such as displacement of the flare-trigger field from the highly sheared PIL and the height of the internal reconnection.
For instance, the internal reconnection in the RS type case may be less sensitive to altitude, and it can occur more frequently than internal reconnection triggered by the OP type structure.
Otherwise the RS type case may be sensitive to displacement of the flare-triggering site from the highly sheared (flaring) PIL.

\begin{figure*}
\epsscale{2.00}
\plotone{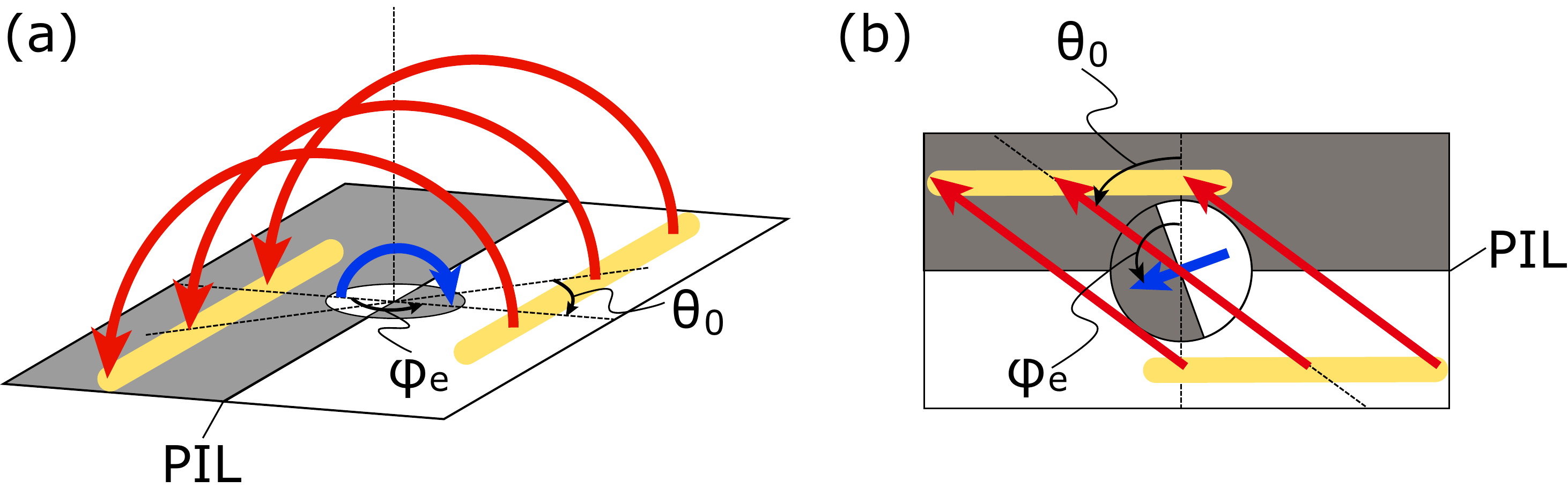}
\caption{
The schematic of the simulation setup in K12 and definition of the angles $\theta_{0}$ and $\varphi_{e}$.
White/black parts represent positive/negative polarity sunspot of an AR.
Red and blue arrows indicate the large-scale sheared field and small-scale bipole field, that can be a trigger of a flare, respectively.
Yellow thick line illustrates two sheared flare-ribbons.
(a) Bird's-eye view. (b) Top view.
}
\label{fig:setting}
\end{figure*}

\begin{figure*}
\epsscale{2.00}
\plotone{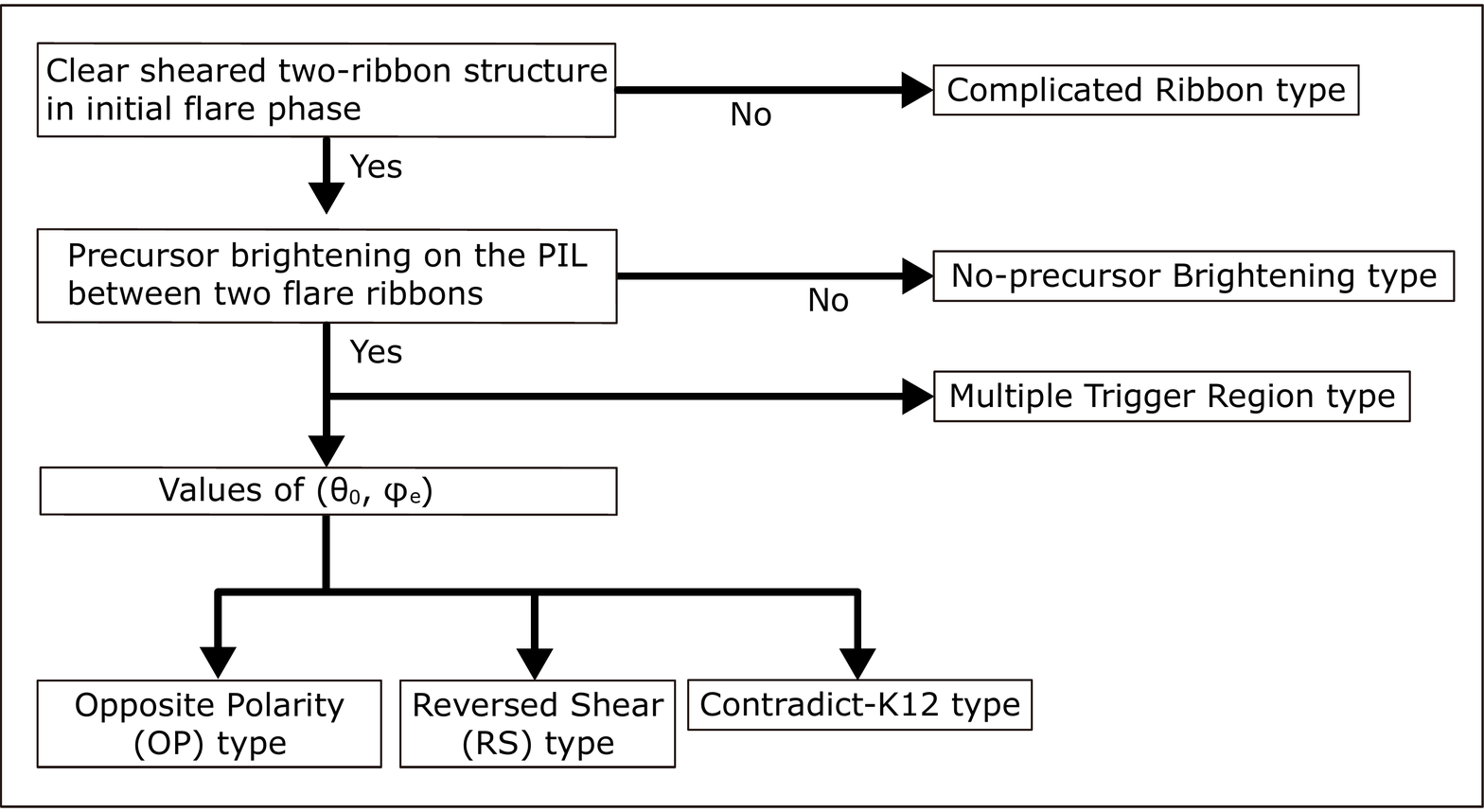}
\caption{
Chart of classification procedure for the selected events.
}
\label{fig:chart}
\end{figure*}

\begin{figure*}
\epsscale{2.20}
\plotone{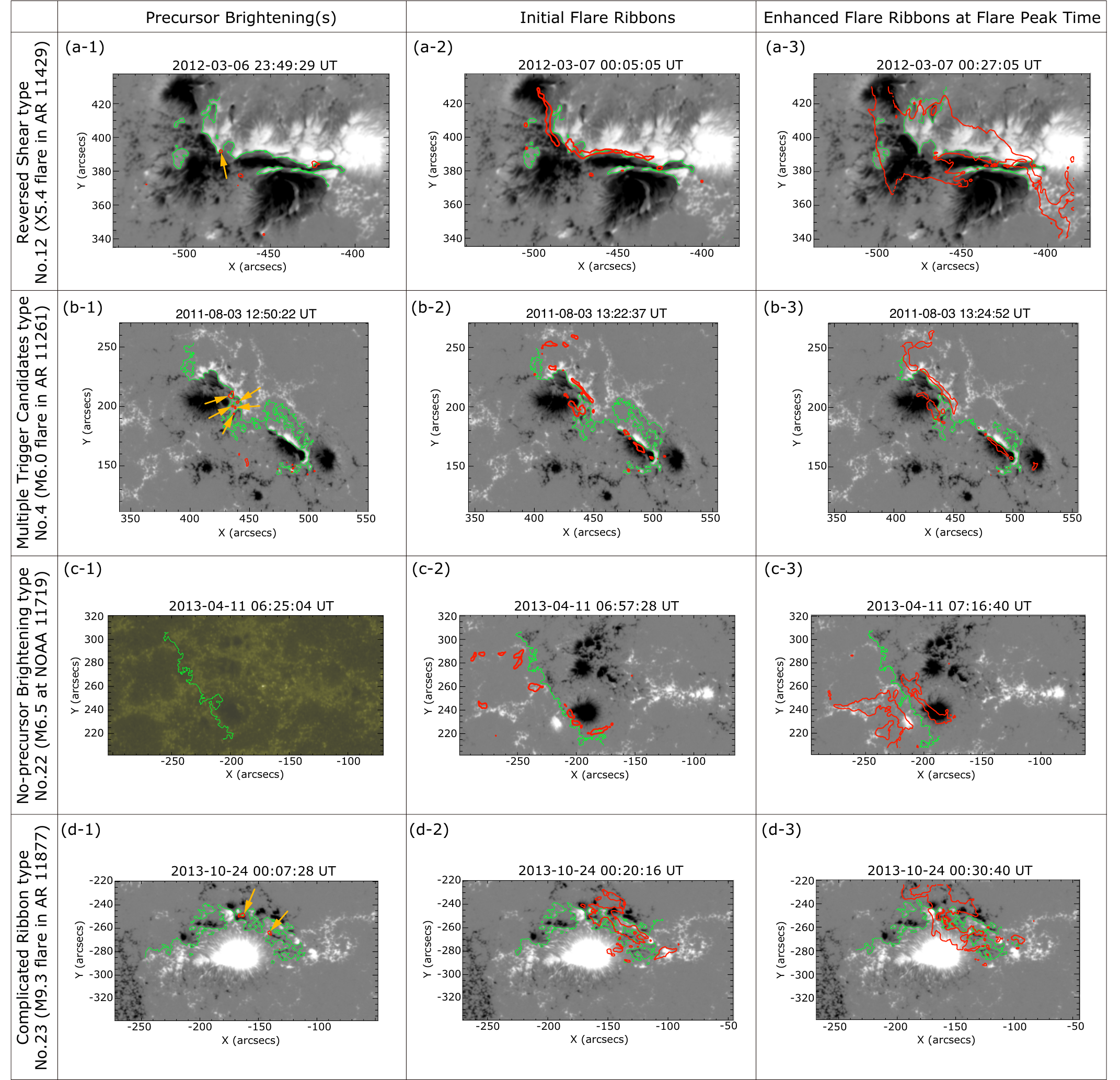}
\caption{
Examples of each event type.
Each line and column are corresponding to the four types and three phases of the flares, as follows. (first line) Reversed Shear type: X5.4 flare in AR 11429 (c.f. Movie 1), (second line) Multiple Trigger Candidates type: M6.0 flare in AR 11261 (c.f. Movie 2), (third line) No-precursor Brightening type: M6.5 flare in AR 11719 (c.f. Movie 3), and (fourth line) Complicated Ribbon type: M9.3 flare in AR 11877 (c.f. Movie 4). (first column) Precursor brightening(s), (second column) initial flare ribbons, and (third column) flare peak time.
Grayscale represents positive and negative polarities of the LOS magnetograms, that are shown in the range $\pm$ 1000 G, and green lines outline the PILs of 0 G.
Red contours plot the initial flare ribbons (2000 DN) seen in AIA 1600 {\AA} images, and yellow arrows in panels (a-1, b-1, and d-1) indicate the precursor brightenings for each event.
Panel (c-1) shoes AIA 1600 {\AA} images 30 minutes before the flare onset because brightenings were not observed on the PIL in the No-precursor Brightening type event.
}
\label{fig:sample}
\end{figure*}

\begin{figure*}
\epsscale{2.00}
\plotone{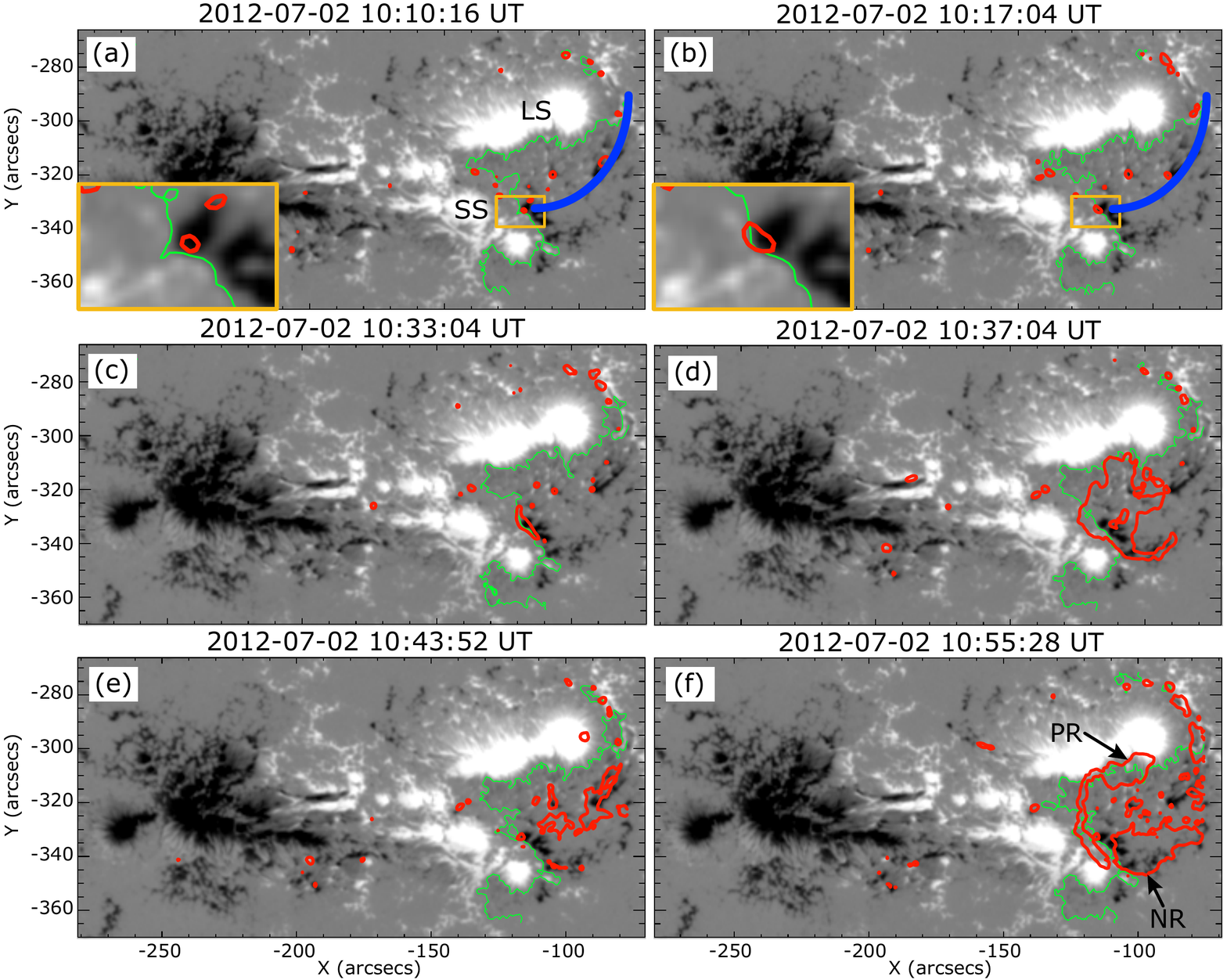}
\caption{
An example of the Complicated Ribbon Type: the M5.6 flare in 2012 July 2 in AR NOAA 11515.
The images are formatted as in Figure~\ref{fig:sample}.
Thick-blue arcs roughly represent the shape and location of a filament that is seen in H$\alpha$ images (c.f \citet{louis14}).
The southern root of the filament is surrounded by yellow rectangles and enlarged in the bottom left in panels (a, b).
A tiny brightening was seen over the PIL within the enlarged region.}
\label{fig:d_11515}
\end{figure*}

In the OP type case, pre-existing large-scale magnetic arcades (red arcades in Figure~\ref{fig:setting}) should directly reconnect with the flare-trigger field (blue arcade in Figure~\ref{fig:setting}), in order to form long twisted flux ropes.
Therefore, the twisted flux rope, which will erupt and trigger flare reconnection, could not be formed if the small bipole is away from the PIL.
Conversely, in the RS type case, magnetic shear cancellation, i.e. the internal magnetic reconnection between pre-existing magnetic field (red arcades in Figure~\ref{fig:setting}) and trigger field (the blue arcade in Figure~\ref{fig:setting}), can occur as long as the small bipole exists under the pre-existing sheared magnetic arcades.
Thus, the sheared magnetic arcades can collapse and flare reconnection can be triggered by shear cancellation as long as the RS type field exists within the sheared arcade even if it is located away from the PIL.
From observational results, such as those reported by \citet{bamba17a}, it is also suggested that the RS type flare-trigger could work even if it is located slightly away from the flaring PIL, even though the trigger field was injected just above the PIL in the K12 simulations.

Therefore, the differences in sensitivity to the geometrical conditions and incidence are originally derived from differences in the physical processes for flare-triggering between the two types.
Thus it is not surprising that the OP type flare-trigger is rare than the RS type flare-trigger.
However, it is still unclear how a distant field that is away from the PIL can trigger flares in the OP and RS cases on the actual solar surface, and we need to statistically investigate the distance between flaring PILs and flare-trigger fields using observational data.

%%%Sec 5%%%
\section{Summary} \label{sec:summary}

In this study, we aimed to evaluate the consistency between the theoretical flare-trigger model of K12 and a variety of major flares that occurred on the actual solar surface, by focusing on the geometrical structure.
We selected 32 major flares and tried to identify the flare-triggering site by applying the analysis method that was developed by \citet{bamba13, bamba14} to SDO/HMI LOS magnetograms and AIA 1600 {\AA} images.
We classified the 32 events into six types of groups including K12's OP and RS types.

The most noteworthy result is that 30\% of the events (including the RS type events and Multiple Trigger Candidates type) were consistent with K12.
Moreover, there was no event that contradicted K12 and satisfied their ``no-flare'' condition at least in the events sampled in this study.
Eight of the 11 Complicated Ribbon type events could be interpreted by K12's flare-trigger model, even though we need more precise analysis of the spatial and temporal relationships between the magnetic field and overlying structures such as filament and precursor brightenings.
Meanwhile, we found that there is a possibility that different physical process(es) from those proposed by K12 cannot also be ruled out for the events which did not show any precursor brightenings, such as the No-precursor Brightening type events and the three Complicated Ribbon type events.
We still can consider other physical processes such as the emerging flux model proposed by \citet{chenshibata00} and the magnetic flux cancellation model proposed by \citet{vanBallegooijen89, Zhang01, Green11}, to trigger these events.
Extended studies are required to reveal the physical process(es) which causes the different types of flares.
Even so, our result that 30\% of the events which were investigated in this study were consistent with the flare-trigger model of K12 leads to the conclusion that the observable features and ($\theta_{0}$, $\varphi_{e}$) parameters of K12 are important to understand the flare triggering.

Moreover, we found that there was no OP type event in the events analyzed in this study while there were six RS type events.
We hypothesized that this difference of incidence between the OP and RS types is likely owing to differences in the physical processes between the two-types: the RS type condition can be satisfied more easily because the geometrical conditions of the RS type are more flexible than those of the OP type.
We need an extended study to clarify the occurrence difference between the OP and RS types, and the consistency/inconsistency between the Complicated Ribbon type, No-precursor Brightening type, and Multiple Trigger Candidates type events and the physical flare-trigger process of K12.
Especially, it is important to comprehensively analyze the relationship between the photospheric magnetic field structures and chromospheric/coronal features, using not only AIA 1600 {\AA} images but also multiple wavelengths.
Moreover, we caution that the limitations of this study given the single UV wavelength used, and relying upon the LOS magnetic field data for the PIL location in AIA images, may introduce bias and errors in the results and their interpretation (see Appendix~\ref{sec:projection}).
As such, it is suggested that further study be performed using additional data and more careful analysis. 
We should reveal the role of the small-scale OP- or RS type field for a flare in various magnetic field topologies, using both simulations and precise observations.

%%%%%%%%%%%%%%%%%%%%%%%%%%%%%%%%%%%%%%%%%%%%%%%%%%%%%%%%%%%%%%%%

%%%ACKNOWLEDGEMENTS%%%
\acknowledgments

The authors deeply appreciate the many instructive comments and encouragements from Dr. K. D. Leka, Dr. Satoshi Inoue, Dr. David H. Brooks, Dr. Shinsuke Imada, and researchers in ISAS/JAXA and NAOJ.
 The authors thank Dr. T. Hara for providing data analysis tool.
The HMI and AIA data have been used courtesy of NASA/SDO and the AIA and HMI science teams.
This work was partly carried out at the Hinode Science Center at NAOJ and Nagoya University, Japan, and the Solar Data Analysis System (SDAS) operated by the Astronomy Data Center in cooperation with the Hinode Science Center of the NAOJ.
This work was supported by MEXT/JSPS KAKENHI Grant Numbers JP16H07478, JP15H05812, JP15K21709, JP15H05814, and JP15J10092.
The authors thank the anonymous referee for valuable comments that improved the clarity of the manuscript.

%スタートアップ/JP16H07478
%新学術総括班/JP15H05812
%国際連携研究/JP15K21709
%A02班/JP15H05814
%学振特別研究員奨励費/JP15J10092

%%%%%%%%%%%%%%%%%%%%%%%%%%%%%%%%%%%%%%%%%%%%%%%%%%%%%%%%%%%%%%%%

%%%APPENDIX%%%
\appendix

\section{Caveat in Using the LOS Magnetic Field Data to Detect Flare-triggering Site} \label{sec:projection}

In this study, we mainly used the HMI LOS magnetograms to investigate the photospheric magnetic field structures.
The LOS magnetic field contains a projection-effect, that is, the LOS component of the magnetic field $\it{B}_{LOS}$ can be treated as the radial component $\it{B}_{r}$ only when the observing angle $\theta$ equal 0 (or $\mu = cos(\theta) = 1.0$).
In other words, $\it{B}_{r}$ can be different from $\it{B}_{LOS}$ depending on the location of the AR on the solar surface, even very close to solar disk center \citep{leka17}.
In this study (also \citet{bamba13, bamba14}), events were selected based on the criterion that the flaring site was located within $\pm~750^{\prime\prime}$ of the solar disk center ($\mu \sim 0.67$), in consideration of this projection effect.

However, we should keep in mind that there is a possibility that we may get a slightly different result between the analysis with $\it{B}_{LOS}$ and with $\it{B}_{r}$.
The flare-trigger field can be a very small-scale structure (empirically estimated as dozens of Mm) compared to the spatial size of an AR, and the spatial size of the trigger field is different between different AR.
Moreover, the azimuth $\varphi_{e}$ is highly sensitive to the configuration of the local PIL and the location of the precursor brightenings in the flare-triggering site.
We should have an extended analysis (a future work) using $\it{B}_{r}$ instead of $\it{B}_{LOS}$ to clarify the consistency between the flare-trigger model of K12 and the events classified as Multiple Trigger Candidates type, No-precursor Brightening type, and Complicated Ribbon type.

We also used SHARP data, converted into Lambert CEA projection, to measure the shear angle $\theta_{0}$ in the present study.
The magnetic field vectors in SHARP data have been transformed into components of the heliographic coordinates ($\it{B}_{r}$, $\it{B}_{\theta}$, $\it{B}_{\phi}$), which were originally in a spherical coordinate system, and these are transformed into a planar CEA coordinate system.
These unit vectors in each coordinate system are not precisely aligned except at the center of the patch.
In fact, an error is caused when the vectors are transformed from a spherical coordinate system to a planar CEA coordinate system (\citet{bobra14}).
In this study, we determined the flaring PIL with co-aligned $\it{B}_{LOS}$ and AIA 1600 {\AA} images, however, it may not mach the PIL identified using SHARP CEA data.
We could have a few-pixel shift in the location of the PIL depending on the location of the AR on the solar surface.
Therefore, we averaged the angle $\theta_{0}$ over a region 5-10 times the area of the flare-trigger field.

%%%%%%%%%%%%%%%%%%%%%%%%%%%%%%%%%%%%%%%%%%%%%%%%%%%%%%%%%%%%%%%%
  
%%% BIBLIOGRAPHY %%%%

%%%References%%%

%%%%%%%%%%%%%%%%%%%%%%%%%%%%%%%%%%%%%%%%%%%%%%%%%%%%%%%%%%%%%%%%

%%%Tables%%%

%Table 1
\begin{table*}
\begin{center}
\begin{tabular}{|c||c|c|c|c|c|c|}
\hline
No. & Date & Onset Time & GOES X-ray & AR & Type & Angle(s) \\
 & & (UT) & Class & (NOAA) & & \\
\hline \hline
	1 & 13-Feb-2011 & 17:28 & M6.6 & 11158 & Reversed Shear & $\theta_{0}\sim88^{\circ}, \varphi_{e}\sim344^{\circ}$ \\ \hline
	2 & 15-Feb-2011 & 01:44 & X2.2 & 11158 & Reversed Shear & $\theta_{0}\sim86^{\circ}, \varphi_{e}\sim331^{\circ}$ \\ \hline
	3 & 09-Mar-2011 & 23:13 & X1.5 & 11166 & Complicated Ribbon & \\ \hline
	4 & 03-Aug-2011 & 13:17 & M6.0 & 11261 & Multiple Trigger Candidates & $\theta_{0}\sim51^{\circ}$ \\  \hline
	5 & 04-Aug-2011 & 03:41 & M9.3 & 11261 & Complicated Ribbon & \\ \hline
	6 & 06-Sep-2011 & 01:35 & M5.3 & 11283 & No-precursor Brightening & $\theta_{0}\sim80^{\circ}$ \\ \hline
	7 & 06-Sep-2011 & 22:12 & X2.1 & 11283 & No-precursor Brightening & $\theta_{0}\sim79^{\circ}$ \\ \hline
	8 & 07-Sep-2011 & 22:32 & X1.8 & 11283 & Complicated Ribbon & \\ \hline
	9 & 08-Sep-2011 & 15:32 & M6.7 & 11283 & Complicated Ribbon & \\ \hline
	10 & 23-Jan-2012 & 03:38 & M8.7 & 11402 & Multiple Trigger Candidates & $\theta_{0}\sim20^{\circ}$ \\ \hline
	11 & 05-Mar-2012 & 02:30 & X1.1 & 11429 & No-precursor Brightening & $\theta_{0}\sim60^{\circ}$ \\ \hline
	12 & 07-Mar-2012 & 00:02 & X5.4 & 11429 & Reversed Shear & $\theta_{0}\sim108^{\circ}, \varphi_{e}\sim313^{\circ}$ \\ \hline
	13 & 07-Mar-2012 & 01:05 & X1.3 & 11429 & Reversed Shear & $\theta_{0}\sim68^{\circ}, \varphi_{e}\sim295^{\circ}$ \\ \hline
	14 & 09-Mar-2012 & 03:22 & M6.3 & 11429 & Reversed Shear & $\theta_{0}\sim74^{\circ}, \varphi_{e}\sim333^{\circ}$ \\ \hline
	15 & 10-Mar-2012 & 17:15 & M8.4 & 11429 & Reversed Shear & $\theta_{0}\sim98^{\circ}, \varphi_{e}\sim312^{\circ}$ \\ \hline
	16 & 10-May-2012 & 04:11 & M5.7 & 11476 & Complicated Ribbon & \\ \hline
	17 & 02-Jul-2012 & 10:43 & M5.6 & 11515 & Complicated Ribbon & \\  \hline
	18 & 04-Jul-2012 & 09:47 & M5.3 & 11515 & Multiple Trigger Candidates & $\theta_{0}\sim56^{\circ}$ \\ \hline
	19 & 05-Jul-2012 & 11:39 & M6.1 & 11515 & Complicated Ribbon & \\ \hline
	20 & 12-Jul-2012 & 15:37 & X1.4 & 11520 & No-precursor Brightening & $\theta_{0}\sim77^{\circ}$ \\ \hline
	21 & 13-Nov-2012 & 01:58 & M6.0 & 11613 & Complicated Ribbon & \\ \hline
	22 & 11-Apl-2013 & 06:55 & M6.5 & 11719 & No-precursor Brightening & $\theta_{0}\sim52^{\circ}$ \\ \hline
	23 & 24-Oct-2013 & 00:21 & M9.3 & 11877 & Complicated Ribbon & \\ \hline
	24 & 01-Nov-2013 & 19:46 & M6.3 & 11884 & No-precursor Brightening & $\theta_{0}\sim82^{\circ}$ \\ \hline
	25 & 03-Nov-2013 & 05:16 & M5.0 & 11884 & No-precursor Brightening & $\theta_{0}\sim79^{\circ}$ \\ \hline
	26 & 05-Nov-2013 & 22:07 & X3.3 & 11890 & Complicated Ribbon & \\ \hline
	27 & 08-Nov-2013 & 04:20 & X1.1 & 11890 & No-precursor Brightening & $\theta_{0}\sim80^{\circ}$ \\ \hline
	28 & 31-Dec-2013 & 21:45 & M6.4 & 11936 & No-precursor Brightening & $\theta_{0}\sim47^{\circ}$ \\ \hline
	29 & 01-Jan-2014 & 18:40 & M9.9 & 11936 & Complicated Ribbon & \\ \hline
	30 & 07-Jan-2014 & 10:07 & M7.2 & 11944 & No-precursor Brightening & $\theta_{0}\sim95^{\circ}$ \\ \hline
	31 & 07-Jan-2014 & 18:04 & X1.2 & 11944 & No-precursor Brightening & $\theta_{0}\sim15^{\circ}$ \\ \hline
	32 & 04-Feb-2014 & 03:57 & M5.2 & 11967 & Multiple Trigger Candidates & $\theta_{0}\sim65^{\circ}$ \\ \hline
\end{tabular}
\caption{Event list}
\label{table:event_list}
\end{center}
\end{table*}

%Table 2
\begin{table*}
\begin{center}
\begin{tabular}{|c|c|}
\hline
Type & Number of Events \\
\hline\hline
Opposite Polarity & 0 \\ \hline
Reversed Shear & 6 \\ \hline
Contradict-K12 & 0 \\ \hline
Multiple Trigger Candidates & 4 \\ \hline
No-precursor Brightening & 11 \\ \hline
Complicated Ribbon & 11 \\ \hline
\end{tabular}
\caption{Summary of the event classification}
\label{table:list_summary}
\end{center}
\end{table*}

%Table 3
\begin{table*}
\begin{center}
\begin{tabular}{|c||c|c|c|c|c|}
\hline
No. & Date & Onset Time (UT) & GOES X-ray Class & NOAA AR & Precursor Brightenings \tablenotemark{a} \\
\hline\hline
3 & 09-Mar-2011 & 23:13 & X1.5 & 11166 & Yes \\ \hline
5 & 04-Aug-2011 & 03:41 & M9.3 & 11261 & Yes \\ \hline
8 & 07-Sep-2011 & 22:32 & X1.8 & 11283 & No \\ \hline
9 & 08-Sep-2011 & 15:32 & M6.7 & 11283 & Yes \\ \hline
16 & 10-May-2012 & 04:11 & M5.7 & 11476 & No \\ \hline
17 & 02-Jul-2012 & 10:43 & M5.6 & 11515 & Yes \\ \hline
19 & 05-Jul-2012 & 11:39 & M6.1 & 11515& Yes \\ \hline
21 & 13-Nov-2012 & 01:58 & M6.0 & 11613 & Yes \\ \hline
23 & 24-Oct-2013 & 00:21 & M9.3 & 11877 & Yes \\ \hline
26 &05-Nov-2013 & 22:07 & X3.3 & 11890 & Yes \\ \hline
29 & 01-Jan-2014& 18:40 & M9.9 & 11936 & No \\ \hline
\end{tabular}
\caption{List of the Complicated Ribbon type events and precursor responses}
\tablenotetext{a}{Precursor brightenings were seen in AIA 1600 {\AA} in at least two frames (i.e. at least over a period of $\sim$1.5~min.) from one hour before to the flare onset.}
\label{table:list_typeD}
\end{center}
\end{table*}

%%%Figures%%%

%Fig 1

%Fig 2

%Fig 3

%Fig 4

%%%%%%%%%%%%%%%%%%%%%%%%%%%%%%%%%%%%%%%%%%%%%%%%%%%%%%%%%%%%%%%%

\end{document}